\documentclass[3p,times]{elsarticle}

\usepackage{ecrc}

\volume{00}

\firstpage{1}

\journalname{Nuclear Physics A}

\runauth{}

\jid{nupha}

\usepackage{amssymb}

\usepackage[figuresright]{rotating}

\begin{document}

\begin{frontmatter}



\dochead{}

\title{Delineating the saturation boundary: linear vs non-linear QCD evolution from HERA data to LHC phenomenology}

\author[label1]{P. Quiroga-Arias}\ead{pquiroga@lpthe.jussieu.fr}
\author[label2,label3]{J.~L. Albacete}
\author[label4,label5]{J.~G. Milhano}
\author[label5]{J. Rojo}

\address[label1]{LPTHE, UPMC Univ. Paris 6 and CNRS UMR7589, Paris, France}
\address[label2]{IPNO, Universit\'e Paris-Sud, CNRS/IN2P3, F-91406, Orsay, France.}
\address[label3]{IPhT, CEA/Saclay, 91191 Gif-sur-Yvette cedex, France.}
\address[label4]{CENTRA, Instituto Superior T\'ecnico, Universidade T\'ecnica de Lisboa, Av. Rovisco Pais, P-1049-001 Lisboa, Portugal.}
\address[label5]{Physics Department, Theory Unit, CERN, CH-1211 Gen\`eve 23, Switzerland.}

\begin{abstract}
The forthcoming p+Pb run at the LHC will provide crucial in formation on the initial state effects of heavy ion collisions and on the gluon saturation phenomena. In turn, most of the saturation inspired phenomenology in heavy ion collisions borrows substantial empiric information from the analysis of e+p data, where abundant high quality data on the small-x kinematic region is available. Indeed, the very precise combined HERA data provides a testing ground in which the relevance of novel QCD regimes, other than the successful linear DGLAP evolution, in small-x inclusive DIS data can be ascertained. We present a study of the dependence of the AAMQS fits, based on the running coupling BK non-linear evolution equations (rcBK), on the fitted dataset. This allows for the identification of the kinematical region where rcBK accurately describes the data, and thus for the determination of its applicability boundary. It also set important constraints to the saturation models used to model the early stages of heavy ion collisions. Finally we compare the rcBK results with NNLO DGLAP fits, obtained with the NNPDF methodology with analogous kinematical cuts. Further, we explore the impact on LHC phenomenology of applying stringent kinematical cuts to the low-x HERA data in a DGLAP fit.
\end{abstract}




\end{frontmatter}


\section{Introduction: situation and strategy}
\label{sec:intro}

The knowledge of the partonic structure of the proton at all relevant observation scales 
plays a crucial role in the analysis of data from present high-energy hadronic colliders, 
most notably at the LHC.
There are different QCD approaches for the description of the scale dependence of parton 
distribution functions. The most commonly used framework are the DGLAP 
equations~\cite{Dokshitzer:1977sg}, 
\begin{equation}
{\partial f(x,Q^2) \over \partial \ln (Q^2/Q_{0}^{2})}=\int_{x}^{1}\frac{dy}{y}   P\left(\alpha_s(Q^2),x/y\right)  f(y,Q^2)\,,
\label{eq:dglap}
\end{equation}
that have been successfully and intensively tested against experimental data. Successful 
as they are, the DGLAP 
 equations are also expected to break down in some kinematic regimes, in particular at 
 small values of Bjorken-$x$.

Analogous resummation schemes aimed at describing the small-$x$ evolution of hadron 
structure, in the direction orthogonal in the kinematic plane to DGLAP evolution, have also
 been developed~\cite{Kuraev:1977fs} (BFKL approach). Additionally, 
 the enhancement of gluon emission at small-x 
 naturally leads to the 
 - empirically observed - presence of large gluon densities and to the need of non-linear 
 recombination terms in order to stabilize the diffusion towards the infrared 
 characteristic of BFKL evolution. Both the resummation of small-$x$ logarithms and the 
 inclusion of non-linear density dependent corrections are consistently accounted for by 
 the B-JIMWLK~\cite{Balitsky:1996ub} equations. 
 Its large-Nc limit, the BK equation, including running coupling corrections (henceforth referred to as rcBK)
 \begin{equation}
\frac{\partial\mathcal{N}(r,x)}{\partial\ln(x_{0}/x)}\!=\!\int d^{2}r_{1}\mathcal{K}(r,r_{1},r_{2})\left[\mathcal{N}(r_{1},x)\!+\!\mathcal{N}(r_{2},x)\!-\!\mathcal{N}(r,x)\!-\!\mathcal{N}(r_{1},x)\mathcal{N}(r_{2},x)\right]\, ,
\label{eq:bk}
\end{equation}
 was shown in~\cite{Albacete:2007yr} to be compatible with experimental data from different
  collision systems (confirmed in~\cite{Albacete:2009fh}).

Based on theoretical arguments alone, one can only strictly establish the applicability of either DGLAP 
or rcBK in their asymptotic limits of very large $Q^2$ or very small $x$ respectively. 
On the phenomenological side, where intermediate ($x$,$Q^2$) kinematics is probed, the situation 
remains unclear. Thus, one needs to define some suitable strategy to identify the regimes
of validity of each formalism and quantify the potential deviation
from these~\cite{Albacete:2012rx}, and this is precisely what we intend to do in the work presented in this proceedings.

The strategy to search for statistically significant deviations from DGLAP evolution was 
laid down in~\cite{Caola:2009iy}, where subsets of data on the reduced DIS 
cross section $\sigma_{r}(x,Q^{2})$ measured at HERA~\cite{:2009wt}, were excluded from 
the fitted data set below some given kinematic cuts 
$Q^{2} \le Q^{2}_{\rm cut} \equiv  A_{\rm cut}\,x^{-\lambda}$, with $\lambda \sim 0.3$ and 
different values of $A_{\rm cut}$, inspired by the generic expectation that possible 
deviations from fixed order DGLAP are larger at small-$x$ and $Q^2$. The PDFs were fitted
 only in the safe kinematical region of the approach, and then backwards DGLAP evolution 
 was used to compare with the excluded, potentialy troublesome, data. The analysis 
 of~\cite{Caola:2009iy} found a systematic discrepancy, albeit with not 
 large enough statistical significance for a decisive statement to be made, indicating 
 that additional dynamics may play a role in 
 the parton evolution in the unfitted region.

\begin{figure}[t]
\begin{center}
\includegraphics[height=7cm]{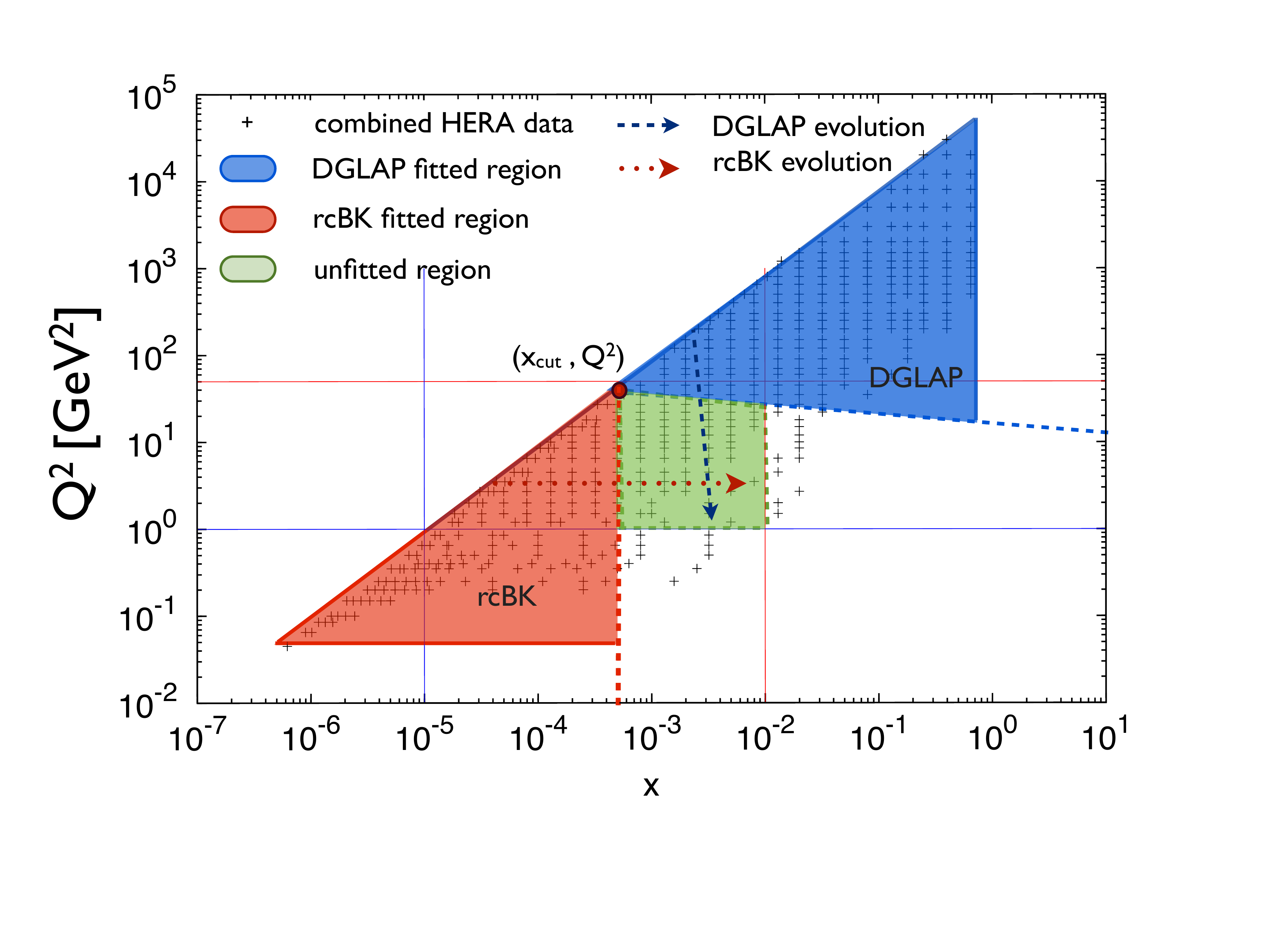}
\end{center}
\vspace{-1.8cm}
\caption{\small Sketch of the kinematic plane with cuts for DGLAP and rcBK fits. The arrows indicate backwards evolution in either formalism to the unfitted {\it test} region. This method provides a direct test of fit stability under changes in the boundary conditions. }
\label{method}
\end{figure}

Following an analogous procedure, we perform fits to data based on the rcBK non-linear evolution equations, limiting the data sets fitted to the safe region of the approach (low-$x$ and $Q^2$), and then study the stability of the fits with respect to the choice of datasets. We systematically reduce the largest experimental value of $x$ included in the fit, $x_{\rm cut}$, and then use the resulting parametrization for the dipole scattering amplitude\footnote{See~\cite{Albacete:2010sy} for a detailed explanation of the AAMQS implementation of the rcBK evolution, and~\cite{Albacete:2012rx} for details on the method.} to predict the value of $\sigma_{r}(x,Q^{2})$ in the unfitted region $x_{\rm cut}<x<x_{0}$. Fig.~\ref{method} summarizes the fitting strategy for the analyses with kinematical cuts.


\section{Results: rcBK (AAMQS) and NNLO DGLAP (NNPDF)}
\label{sec:results}
\begin{figure}[t]
\begin{center}
\vspace{-0.3cm}
\includegraphics[height=6cm]{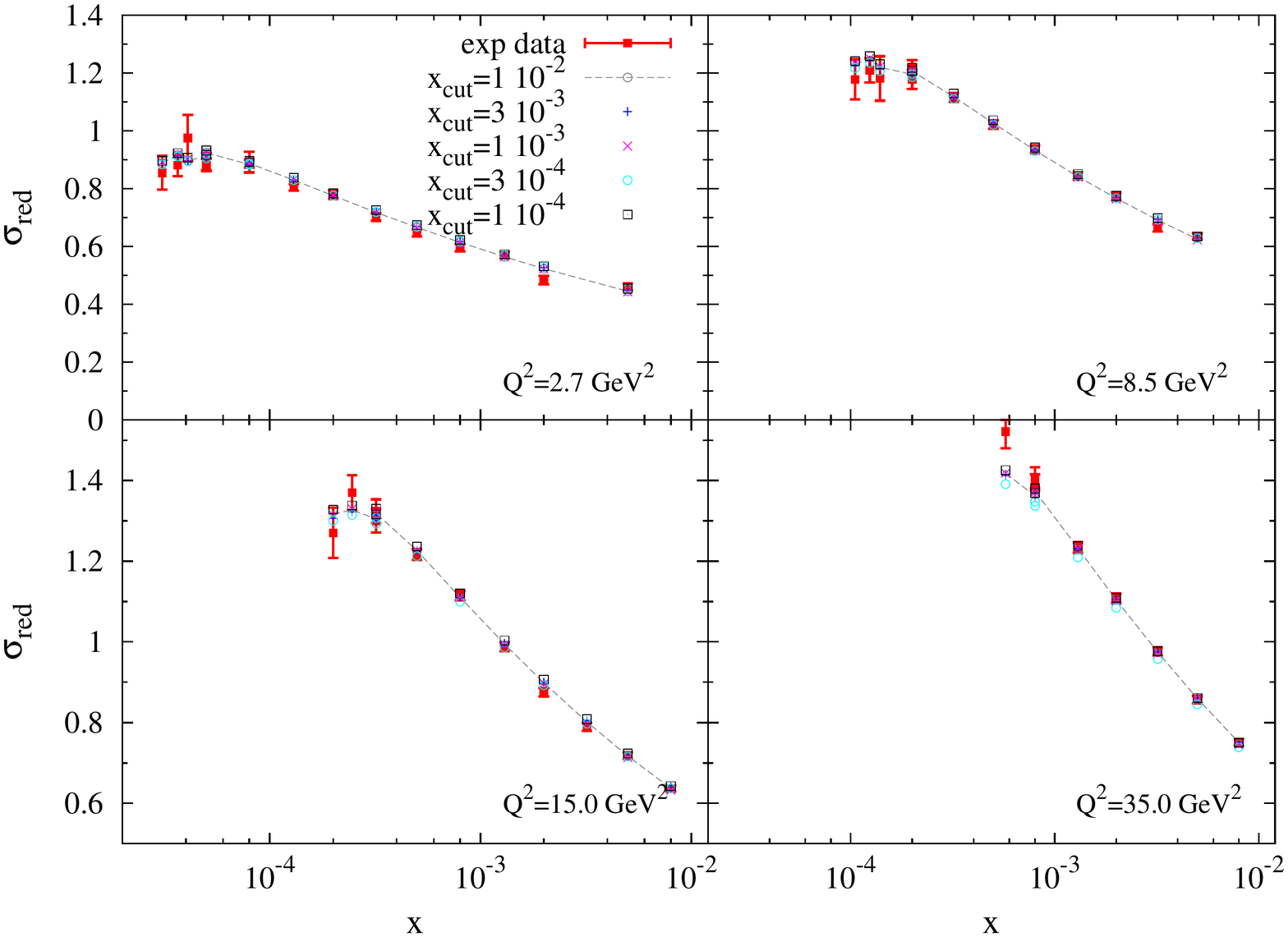}
\includegraphics[height=6cm]{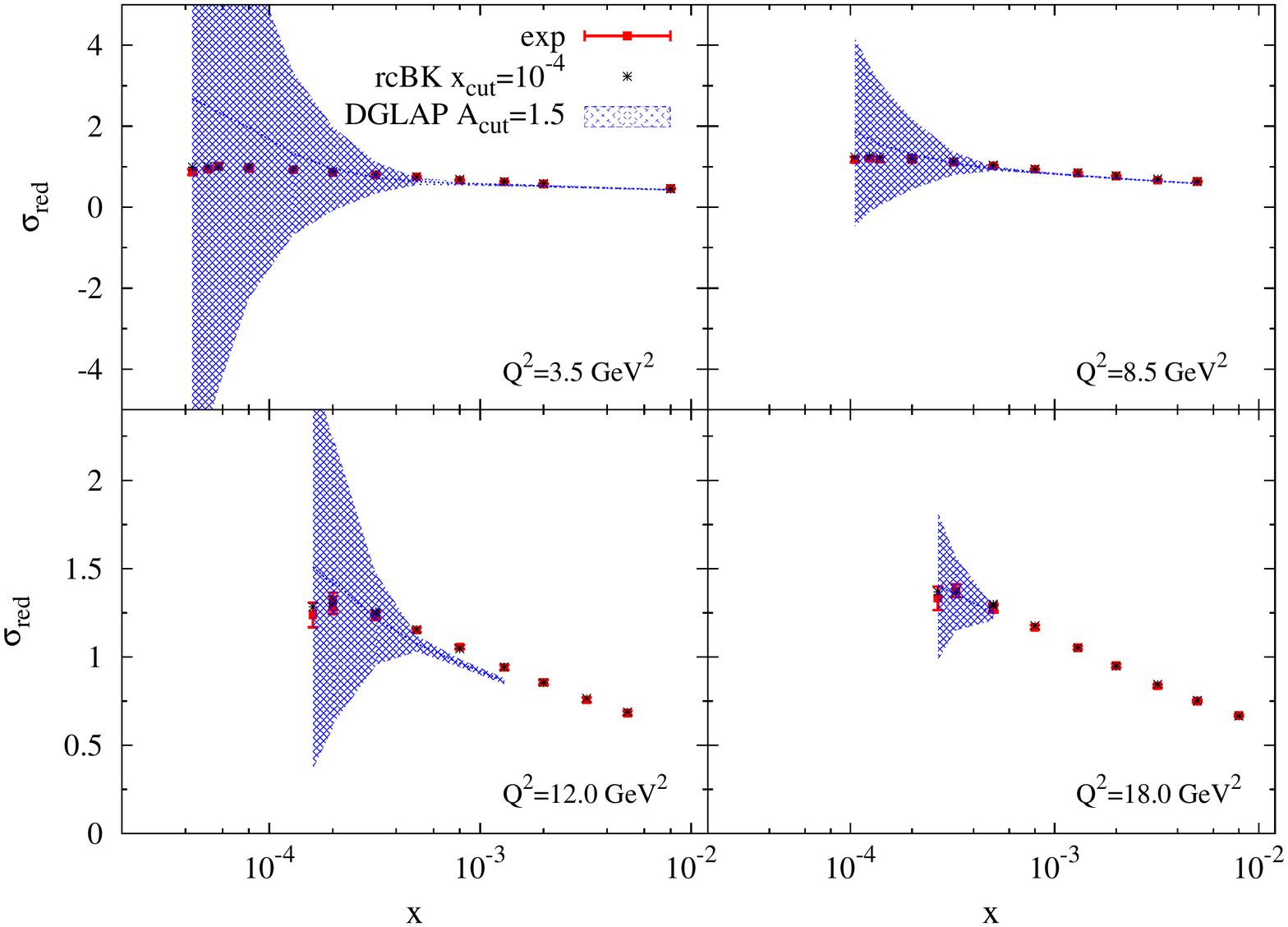}
\end{center}
\vspace{-0.5cm}
\caption{(\emph{left}) Comparison of the result for reduced cross section obtained with rcBK fits with different cuts
and HERA data for four different bins in $Q^{2}$. (\emph{right}) rcBK cut fit with $x_{\rm cut}=10^{-4}$ and the DGLAP fit with $A_{\rm cut}=1.5$, compared to
the experimental HERA-I data. The comparison is shown
in four different bins in $Q^{2}$. In the DGLAP case the band corresponds to the PDF uncertainties.}
\label{rcBKdata}
\end{figure}

We now show the results with various kinematical cuts
obtained with rcBK and DGLAP evolution equations. Fig.~\ref{rcBKdata}-left shows the 
comparison of the theoretical results stemming from rcBK fits to data with different 
$x$-cuts from $x_{\rm cut}=10^{-2}$ to $10^{-4}$. The quality of the fits is comparably 
good independently of the cut, despite the decreasing number of points with decreasing 
$x_{\rm cut}$. Also the extrapolations of the results for $\sigma_r$ from fits with cuts 
to the unfitted region , i.e to $x>x_{\rm cut}$, yield a good description of the data. 
Fig.~\ref{rcBKdata}-right shows the results corresponding to the rcBK fit with the most 
stringent cut, $x_{\rm cut}=10^{-4}$, together with experimental data and the analogous 
results from the NNLO DGLAP fit with cut $A_{\rm cut}=1.5$. While the DGLAP extrapolations 
to the unfitted, {\it test} region are compatible with data within the uncertainty bands, 
the central values of the predictions show significant deviations from data in the region 
of small-$x$. 

\begin{figure}[t]
\begin{center}
\includegraphics[height=4.5cm]{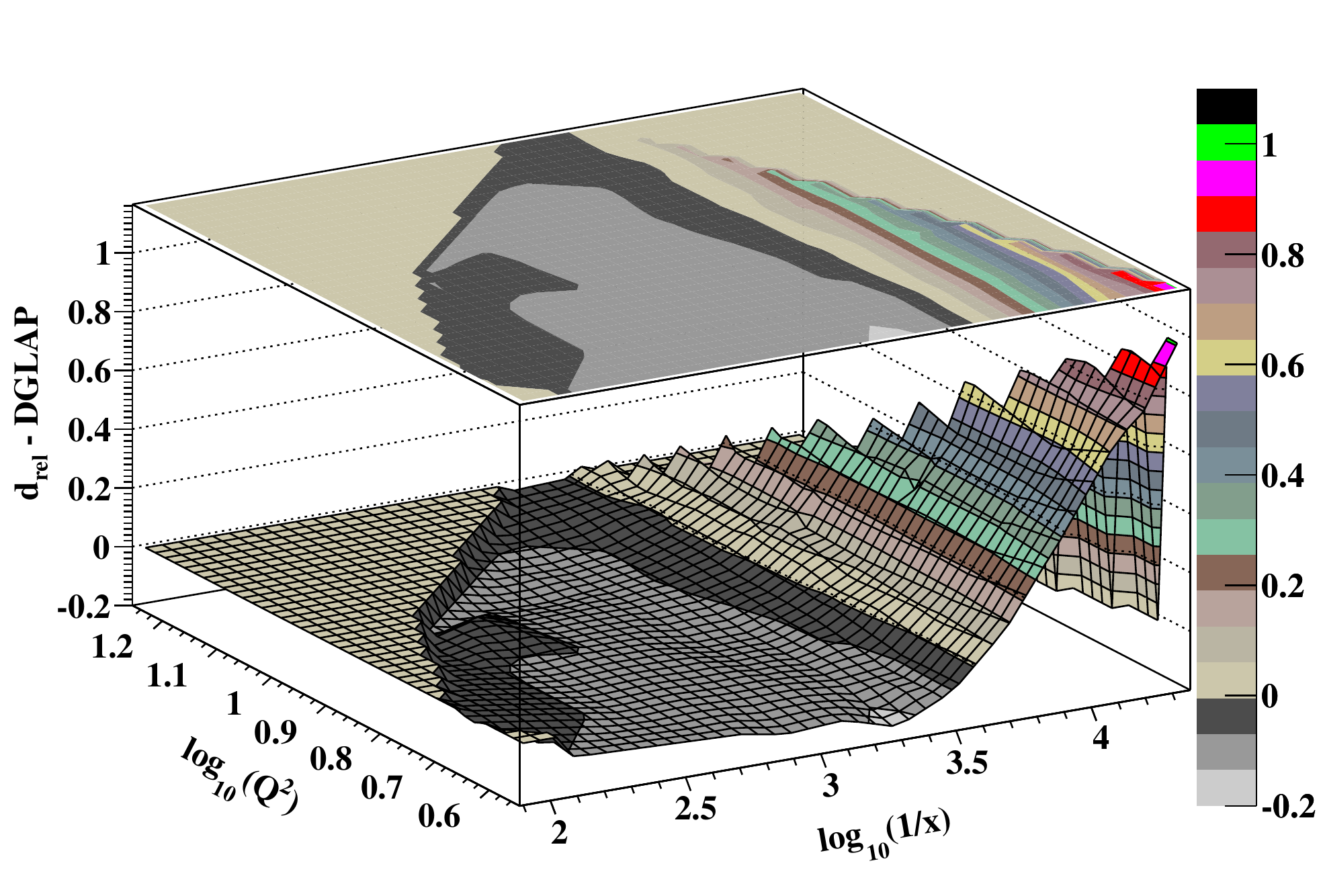}
\includegraphics[height=4.5cm]{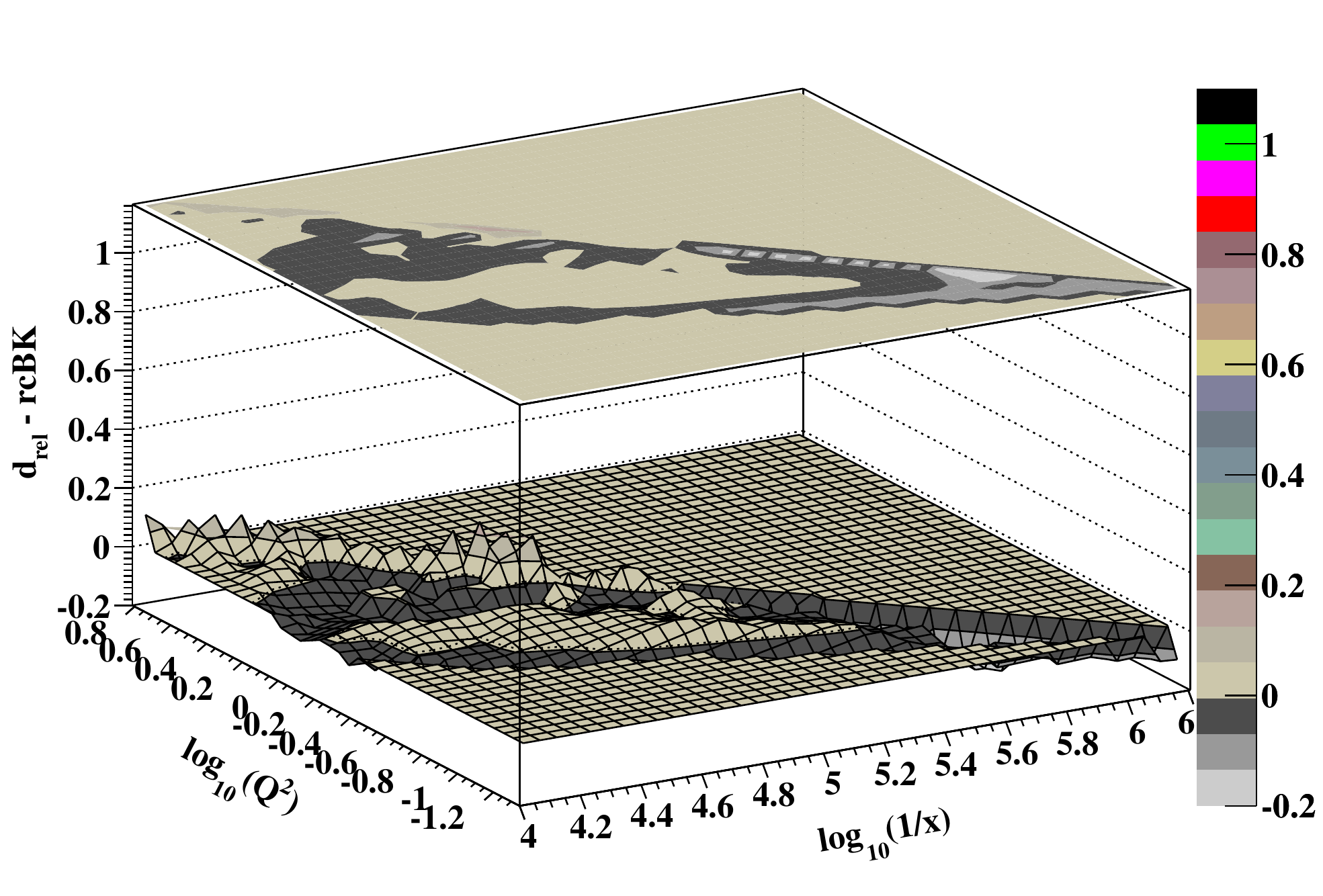}
\end{center}
\vspace{-0.7cm}
\caption{\small The relative distance, $d_{\rm rel}(x,Q^{2})$, for DGLAP (left) and rcBK (right) cut fits.}
\label{drel}
\end{figure}

We quantify these deviations by calculating 
the relative distance between the theoretical results and experimental data, 
$d_{\rm rel}(x,Q^{2})=\frac{\sigma_{\rm r,th}-\sigma_{\rm r,exp}}{(\sigma_{\rm r,th}+\sigma_{\rm r,exp})/2}\,$,
 both for the 
 rcBK and DGLAP cut fits, with cut values $x_{\rm cut}=10^{-4}$ and $A_{\rm cut}=1.5$ 
 respectively. As shown in Fig.~\ref{drel}, $d_{\rm rel}$ is on average much smaller for the 
rcBK fits than it is for the DGLAP one, the latter also showing a systematic trend to 
underestimate data at small-$x$ and to overshoot them at larger $x$. In turn, the rcBK 
values for $d_{\rm rel}$ alternate in sign in all the unfitted region.

\begin{figure}[t]
\begin{center}
\includegraphics[height=6.1cm]{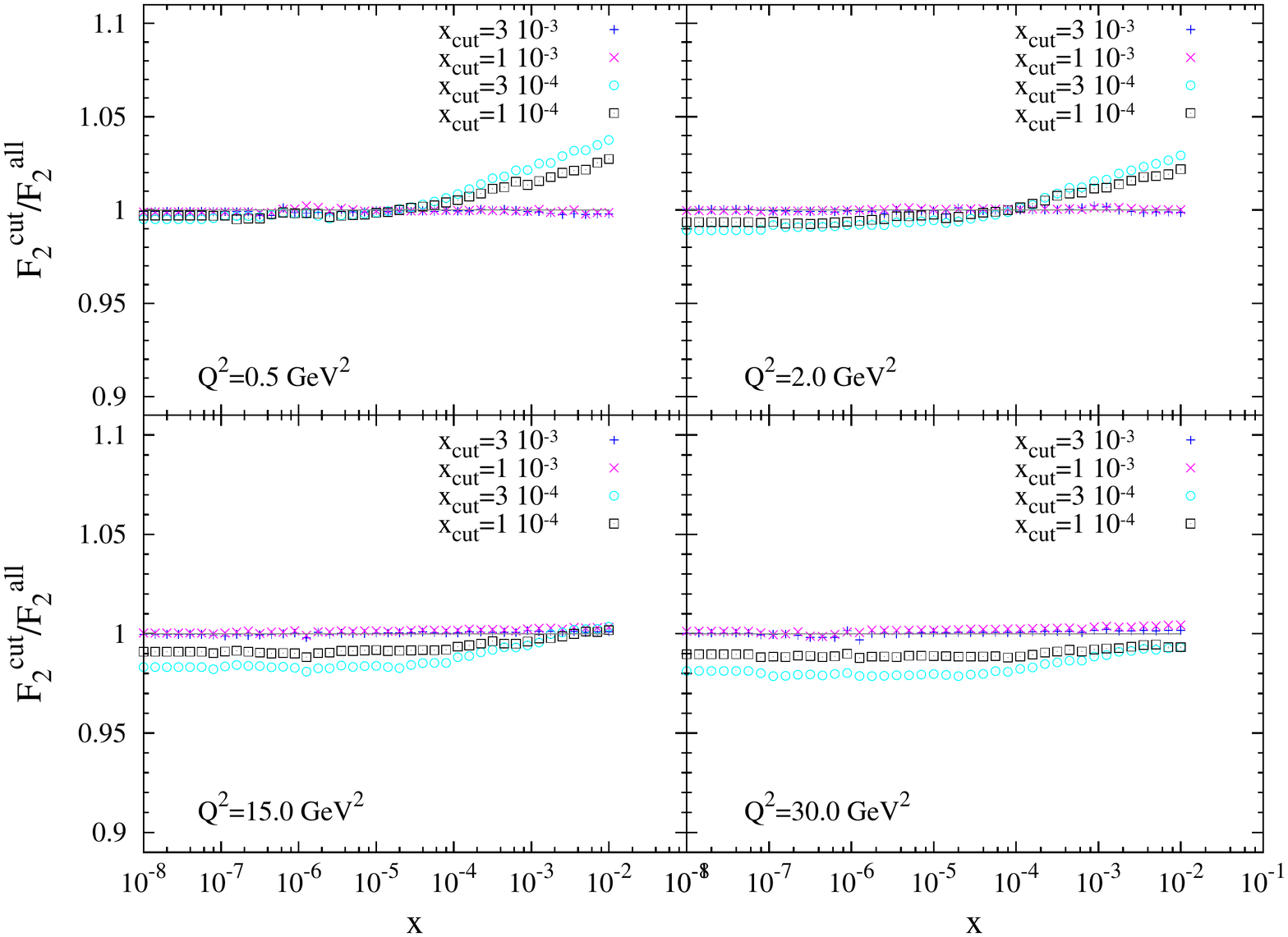}
\includegraphics[height=6.1cm]{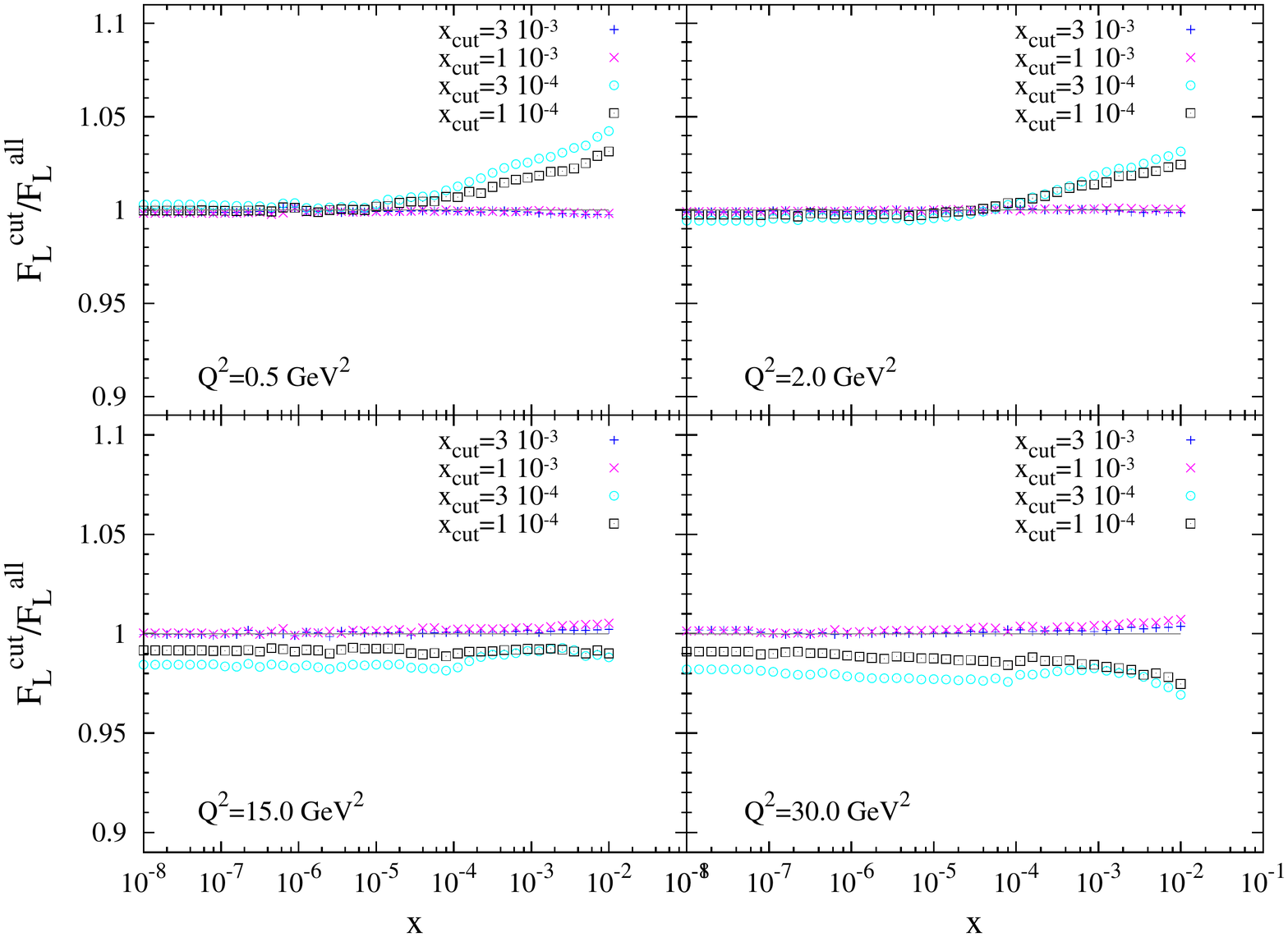}
\end{center}
\vspace{-0.8cm}
\caption{\small Extrapolation to the low-x  region from the rcBK cut fits presented in Fig.~\ref{rcBKdata}. The total, $F_2(x,Q^2)$ (\emph{left}), and longitufinal, $F_L(x,Q^2)$ (\emph{right}),  structure functions are calculated down to $x=10^{-8}$. The results are presented as a ratio of the prediction for the different cut fits to the prediction for the uncut fit, i.e. a fit to all data with $x<x_0=10^{-2}$ and $Q^2<50$ GeV$^2$.}
\label{fig:lowx-extrapol}
\end{figure}

In order to explore the predictive power of the rcBK approach and the sensitivity to boundary 
effects encoded in the different initial conditions for the evolution under the inclusion/exclusion 
of subsets of data we extrapolate our results for the total $F_{2}(x,Q^{2})$ and longitudinal  
$F_{L}(x,Q^{2})$ structure functions to values of $x$ smaller than those currently available 
experimentally. The results, Fig.~\ref{fig:lowx-extrapol},  show that the predictions stemming 
from different fits converge, within approximately one percent accuracy, at values of $x\sim10^{-4}$. 
These predictions could be verified in planned facilities as the LHeC~\cite{lhec2} or the 
EIC~\cite{EICwhite}, where a much extended kinematic reach in $x$ would be 
available

\begin{figure}[t]
\begin{center}
\includegraphics[height=4.4cm]{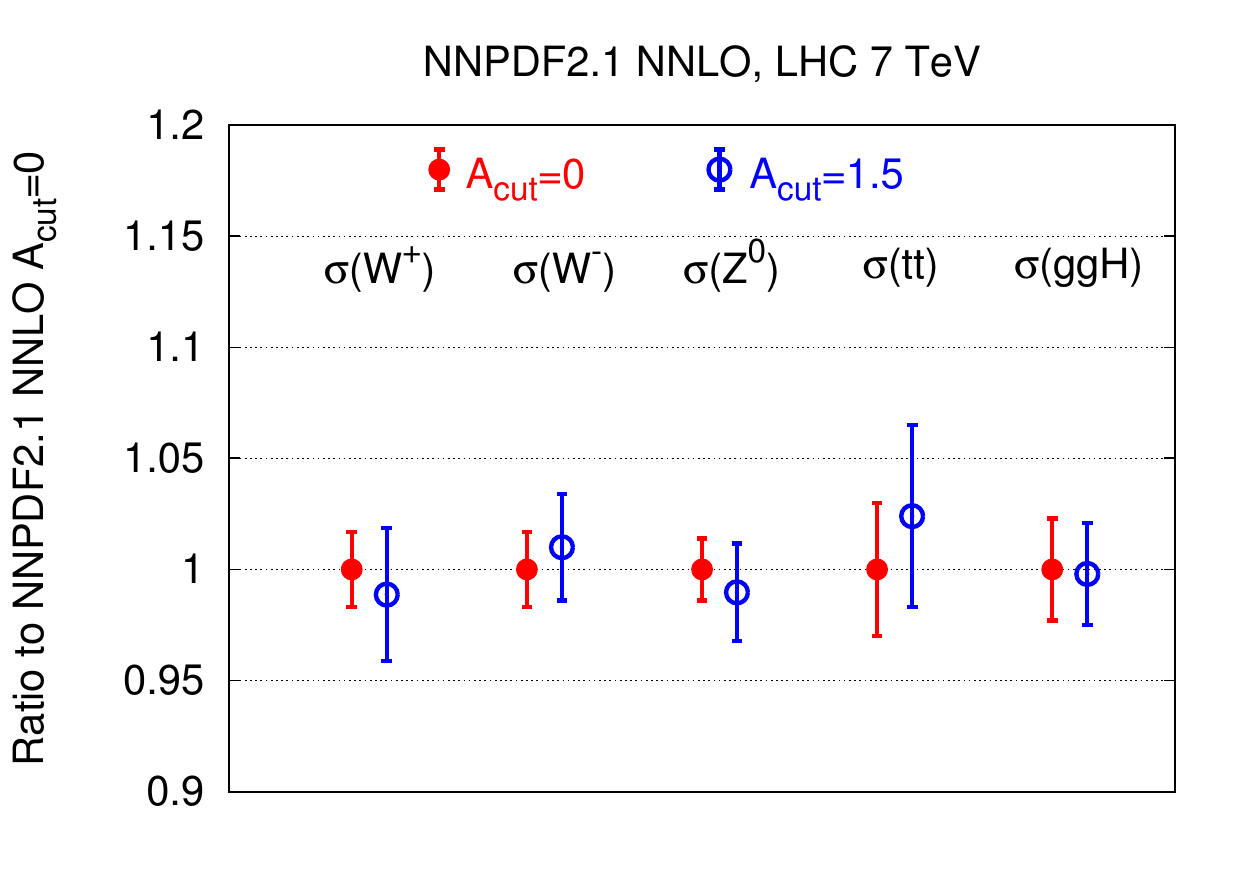}
\includegraphics[height=4.4cm]{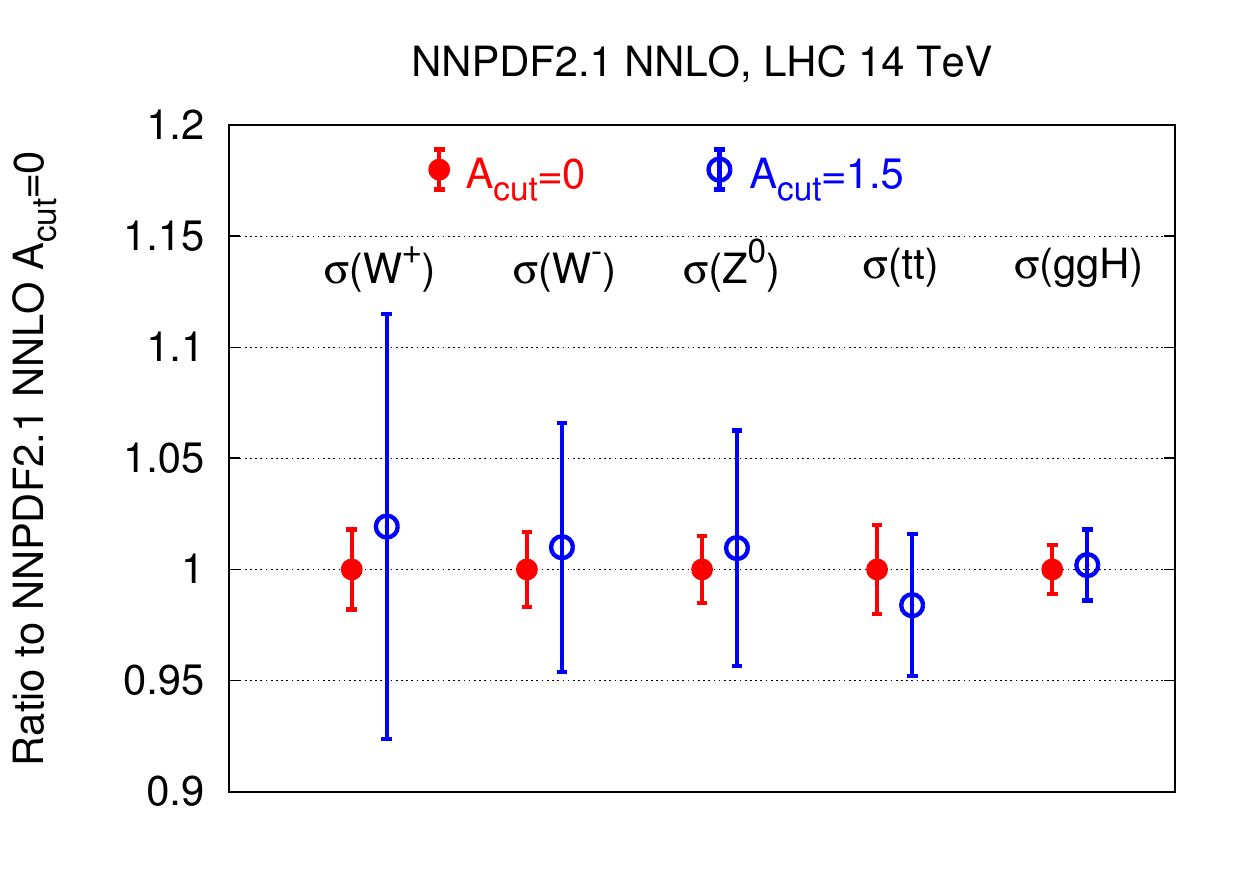}
\end{center}
\vspace{-0.8cm}
\caption{\small Comparison of the predictions for LHC NNLO cross sections
for the reference NNPDF2.1 NNLO fit with $A_{\rm cut}=0$ and
with the NNPDF2.1 NNLO fit with $A_{\rm cut}=1.5$. Cross sections are shown
as ratios to the uncut $A_{\rm cut}=0$ predictions. We show results
both for LHC 7 TeV (left plot) and for LHC 14 TeV (right plot).
\label{fig:pheno}}
\end{figure}

To conclude, we need to explore the impact that potential deviations from DGLAP evolution may have on LHC phenomenology. We compute benchmark LHC cross sections with the PDF sets both with and without the small-$x$ kinematical cuts using the NNPDF2.1 NNLO set. The results are shown in Fig.~\ref{fig:pheno}. While the impact of cutting the
small-$x$ and small-$Q^2$ HERA data from the fit is rather moderate at
LHC 7 TeV, at LHC 14 TeV the effect is much larger, since smaller values of $x$ in the PDFs are being probed. 
One can observe that the cross section for Higgs boson production in gluon fusion is very stable against the kinematical cuts, while for the electroweak boson and top production cross sections the PDF uncertainties
increase by up to a factor five. This needs to be carefully considered, since these processes constitute an important background in Higgs searches.

\bibliographystyle{elsarticle-num}
\bibliography{<your-bib-database>}



\end{document}